\documentstyle[12pt]{article}

\begin{document}

\topmargin 0pt
\oddsidemargin 5mm

\setcounter{page}{1}
\vspace{2cm}
\begin{center}

{\bf STIMULATED RADIATION OF PARTICLES IN CRYSTALLIC UNDULATORS }\\
\vspace{3mm}
{\large  R.O.Avakian, L.A.Gevorgian, 
K.A.Ispirian$^{*)} $ and R.K.Ispirian }\\
\vspace{3mm}
{\em Yerevan Physics Institute, Brothers Alikhanian 2, Yerevan, 375036, Armenia}\\
\end{center}

\vspace{3mm}
\centerline{\bf{Abstract}}
It is investigated the stimulated radiation which arises when the 
density and energy of the beam of relativistic particles channeled in 
microscopic crystallic undulators (CU) obtained by the application 
of transverse ultrasonic (US) oscillations to single crystals, exceed certain 
values. For positron beams expected at future linear colliders and real CU
it is given the results of numerical calculations for spontaneous and stimulated 
radiation. 

\indent
PACS61.80 MK

\vspace{10mm}

\indent
The problems connected with the construction of x-ray and gamma lasers remain 
still unresolved (see [1]). The possibility of the production of the stimulated 
channeling radiation has appeared to be almost hopeless since it requires 
particle beams of very high density for which the crystals will be damaged (see 
[2]). Nevertheless, arrangements are under construction at present for 
production of high brightness x-ray beams in which the self amplification of the 
spontaneous emission (SASE, the operation principles of which are proven for 
cantimeter and micron wavelenths) will take place (see [3]).

\indent
Looking for intense radiation sources in the works [4,5] it has been considered 
the radiation produced by positively charged particles in CU. The influence of 
the medium polarization on the spontaneous radiation produced in CU has been 
taken into account in [6]. The estimates [5] show that at high positron beam
densities this radiation becomes stimulated. Taking into account the above 
practical interest this work is devoted to the investigation of various 
characteristics of the CU stimulated radiation taking into account 

\indent
--------------------------------------------------------

\indent
*) e-mail:ispirian@vx1.yerphi.am

\indent
the beam and medium density effects. 

\indent
Using the formula (4) of the 
work [6] after necessary folding one can show
that for a positron beam with gaussian distribution of the energy $ E $ or 
of the Lorentz factor $ \gamma = E/mc^2 $ of the particles arround 
$ \overline{\gamma}= \overline{E}/mc^2 $ with dispersion $\sigma $, the 
spectral distribution of the number of photons of 
the spontaneous radiation produced by a channeled particle in a CU with length 
$ L $ has the form
\begin{equation}
\label{AA}
\langle \frac{dN}{d\xi }\rangle = D f(\xi) F[X(\xi )],
\end{equation}
where 
$$
f(\xi) = \frac{1}{2} \left[ [ (1+\eta^2-\eta^2_c )\xi + \frac{1}{\xi} 
- 1]^2 + 1 \right],
$$
$$
F[X(\xi )] = 1 - erf[X(\xi )],
$$
$$
X(\xi ) = \frac {1}{\sqrt{2} \sigma } \left[ \frac{\xi}{p \eta \sqrt 
{(\xi -\xi_1)(\xi_2 - \xi)}} - 1 \right],
$$
$$
\xi_{1,2} = \frac {1}{1 \pm \sqrt{1-\eta^2_c}},
$$
$$
\eta_c = \sqrt{1-(\gamma_0^2 / \overline{\gamma})^2},
$$

\indent
$ D = \pi \alpha \eta^2 L/l $, $ \alpha = 1/137 $. As in [6] $ \eta = 
\sqrt{2} \pi A / \lambda_p $, $ \lambda_p = 2\pi c/ \omega_p $ plasma  
wavelength of the CU, $ \gamma_0 = \omega_p/\Omega_p =l/\lambda_p $, $ \Omega_p 
= 2 \pi c/l $, $ \lambda_0 = l/\gamma_0 ^2 = \lambda_p^2 /l $ is the wavelength
arround which the narrowing of the spectrum takes place due to the polarization 
of the medium [6], while $ A $ and $ l = v_{us}
/ f_{us} = 2 \pi c / \Omega $ are the amplitude and the period of the CU
obtained as a result of application of transversal US oscillations with
 $ f_{us}$ and velocity $v_{us}$. Instead of x and k of the work [6] it is 
introduced new dimensionless energy of photon $ \xi = \omega/ \Omega \gamma_0^2 = 
\lambda_0 / \lambda $ and positron $ p = \overline{ \gamma} / \gamma_0 = 
1/\sqrt{1-\eta_c^2} $. The radiation takes place when $ \gamma \geq \gamma_
{thr} = \gamma_0 / \sqrt{1-\eta^2} $ or $ p \geq p_{thr} = 1 / \sqrt{1-
\eta^2 } $.

\indent
The spontaneous radiation spectral distributions of positron beams, calculated 
with the help of the formula (1), differ from the corresponding curves of ref.
[6] by their smooth boundaries due to the energy spread of the beam particles.

\indent
In the validity limits of the one dimensional (1D) FEL theory in which the 
diffraction is not taken into account, following the works [3,7] and 
neglecting the angular divergence of the beam with density $ n $, it can shown
that for low gains of the stimulated radiation when the gain $ G\leq 1 $ one 
has
\begin{equation}
\label{AB}
G(\xi) = G_0 \frac{p}{ \sigma} \varphi (\eta ) \psi[X(\xi)],
\end{equation}
where
$$
G_0 = \frac{(2\pi)^3 r_0 n L^2}{\gamma_0^3},
$$
$$
\varphi (\eta ) = \frac{\eta^2 (1-\eta_c^2 + \eta^2 )^2 }
{ 1 + \sqrt{ \eta_c^2 - \eta^2}},
$$
$$
\psi[X(\xi)] = exp[-X^2(\xi )] F[X(\xi )],
$$
$ r_0 = e^2/mc^2 $ is the electron classical radius.

\indent
For $\eta \rightarrow 0 $ the function $ \varphi (\eta ) $, and therefore $ G $ 
goes to 0,while for $p \gg 1 $ and $\eta 
\rightarrow \eta_c $ the function $\varphi(\eta )\rightarrow 1 $, and $ G $ 
achievs its maximal value. The function $ \psi(\xi) $, therefore  $ G $ too 
have their maximum when
\begin{equation}
\label{AC}
\xi_0 = \frac{1}{1-\sqrt{\eta_c^2 -\eta^2}}
\end{equation}
with an half width of the stimulated radiation spectral distribution
\begin{equation}
\label{AD}
\Delta \xi \approx \sqrt{1 - \eta^2 }\xi_0^2 \sigma.
\end{equation}

\indent
The maximal value of $ G_{max} $ is obtained at $ \eta \rightarrow \eta_c $ 
( $ \xi_0 = 1$, $ \Delta \xi = \sigma / p $ ) and $ p\gg 1 $:
\begin{equation}
\label{AE}
G_{max} \approx \frac{p}{ \sigma} G_0 = n/n_c,
\end{equation}

\begin{equation}
\label{AF}
n_c = \frac{\sigma \gamma_0^3}{(2\pi )^3 r_0 p L^2}=
\frac{ \gamma_0^4}{(2\pi )^3 r_0  L^2 \overline{\gamma}},
\end{equation}

\indent
where $ n_c $ is the beam density for which the gain 
is of the order of 1. The further increase of the density results in 
the growth of the stimulated radiation intensity.

\indent
When $ \sigma $ decreases, $ \Delta \xi $ also decreases and the values of 
$ G $ and $ G_{max} $ increase. Let us note that in contrast to the usual  
FELs without filling medium $ \gamma_0 $ enters into the expressions for 
$ G_0 $ and $ n_c $ instead of $ \gamma $ that results in essential increase 
of the gain as in the case of gas loaded FELs [8].

\indent
In the nearest future it will be obtained beams with  $ \sigma \leq 
0.001 $ and particle density $ n = 1.0.10^{21} cm^{-3} $ at $ \overline{E} 
= 250 $ and 50 GeV in the interaction point of the $e^+e^- $ collider TESLA [9];
$ n = 1.8. 10^{18} cm^{-3} $ at $ \overline{E} = 10- 25 $ GeV in the TESLA x-ray 
FELL undulator with $ L = 87 m $ and $ n = 6.4.10^{16}$ at $\overline{E}$ 
= 1 GeV in the TTF TESLA undulator with $ L = 27 m $ [10]. Let us note that at 
$\overline {E} $ = 1 and 10-25 GeV the beams will be electron beams. However,
corresponding positron beams can be produced after the TESLA launching, and
the beam densities can be increased by 1-2 orders with the help of focusing 
because the lenth of CU is much shorter than the length of SASE FELs.

\indent
The necessary parameters and results on the stimulated radiation calculated 
for these beams with the help of the formulae (2)-(6) for quartz CU (see [6]) 
are given in Table. After the values of the energy (first column) the second 
column gives the the CU thicknesses which are permitted by dechanneling and 
other technical reasons. The third column shows the desired maximal numbers 
of the CU periods. These values of $E $, $L $ and N determine $ f_{us}$ 
(fourth column),$ l $, $ \Omega $, $ \gamma_0 $, $ p $, $ \eta_c $ and $ n_c $
(fifth column). The sixth column gives the choosen values of the US 
amplitudes which provide the necessary values of  $ \eta $ и $ \xi_0 $ , 
i.e. the stimulated radiation photon energies  $ \omega_0 = \xi_0 \Omega 
\gamma_0^2 $ (seventh column). In the next two columns it is given 
the relative half widths $ \Delta \omega /\omega = \Delta \xi / \xi_0 $ 
and the gains calculated with the help of the formula (2). As it follows 
from the Table  besides the case $ \overline{E} = 1 $ GeV, 
even for densities $ n \approx n_c \approx 10^{13} см^{-3} $, 
much less than the ones of the future beams one, can obtain values of 
$ G $ of the order of 1. It is clear that values $ G \gg 1 $
can be obtained with the help of beams with higher density, and intense 
monochromatic beams can be produced with photon energies
$ \omega_0 = 33 $; 55 and 100 keV,i.e. with energies much higher than 
the photon energies which will be produced at the linear accelerators 
SLAC and TESLA.   

\indent
The intensity of the radiation from CU per beam particle is equal to the 
product of (G+1) and spontaneous radiation spectrum (1). It is clear that 
for  $ G \leq 0.2 $ there is little difference between the spectra of the 
spontaneous and stimulated radiations at the CU exit. Therefore, for small 
values of $ G $ the stimulated radiation in CU will be of interest if 
x-ray cavities (see, for instance [1]) will be created in which the losses 
per one cycle is less than $ G $, and there is radiation photon storing from 
the consequent bunches of positron pulses. The photons produced and stored 
in the cavities by the bunches of positron beam macropulses can serve as 
"X-ray target" for the study of many processes [11].

\indent
The gain and the spectra of the stimulated radiation for high 
gain FELs with $ G\gg 1 $ are considered within the 
1D FEL theory in many works (see [3, 12, 13]). In conclusion
without considering these and other problems connected with the multiple 
scattering ( the CU thickness is less than the dechanneling length, so 
that the particles remain channeled), with 
the US amplitude and frequency spread, with diffraction and with particle 
density modulation and beaming when $ G \gg 1 $, let us estimate the 
possibility of the SASE process in CU. The Pierce parameter and the length 
for e time enhancement of the intensity of SASE FELs can be written in the 
form  $ \rho = 2.82.10^{-5} (K^2l^2n)^{1/3}$ and 
$ L_g = 3.26.10^3 (K^{-2}ln^{-1})^{1/3} $, respectively, where $ K=2 \pi \gamma 
A /l $. The values of $ \rho $ and $ L_g $ for the above considered CU and beams 
with the expected densities are given in the last columns of the Table, 
respectively. As it follows from such a direct application of the SASE 
theory stimulated radiation with intensity exponential growth and saturation 
processes can also take place in CU. In future the use of nanotubes with 
much smaller $ l $ and larger dechanneling lengths [14] can make easier 
the problem of the production of SASE sources of intense x-ray  beams.

\newpage

Table \\
\begin{tabular}{|c|c|c|c|c|c|c|c|c|c|c|}
\hline
E& L& N& f& $n_c10^{-13} $& A& $\omega_0 $&
$\Delta \omega / \omega $& G& $\rho 10^3 $& $L_g$ \\
\hline
GeV& cm& -& MHz& $cm^{-3}$& nm& keV& $\%$& -& -& cm\\
\hline
250& 10& 1000& 57& 1.5& 8.35& 101& 0.01& 0.68& 1& 0.85 \\
\hline
50& 5& 1000& 114& 1.9& 8.26& 54.7& 0.02& 0.51& 2& 0.25 \\
\hline
25& 2.5& 1000& 228& 0.94& 7.94& 33.1& 0.05& 0.7& 0.3& 0.84 \\
\hline
1& 0.2& 200& 570& 93.4& 7.8& 13.6& 0.06& 0.007& 0.3& 0.35 \\
\hline

\end{tabular}

\end{document}